\def\Re{{\text{Re}}\,}
\def\epsilonF{\epsilon_{\text{F}}}

\def\vF{v_{\text{F}}}
\def\NF{N_{\text{F}}}

\def\qslash{q\!\!\!/}
\def\be{\begin{equation}}
\def\ee{\end{equation}}
\def\bea{\begin{eqnarray}}
\def\eea{\end{eqnarray}}
\def\bse{\begin{subequations}}
\def\ese{\end{subequations}}

\documentclass[prl,twocolumn,showpacs,preprintnumbers,amsmath,amssymb]{revtex4}
\usepackage{graphicx}
\usepackage{dcolumn}
\usepackage{bm}
\begin{document}
\preprint{}
\title{Anomalous Density-of-States Fluctuations in Two-Dimensional Clean Metals}
\author{T.R. Kirkpatrick$^{1}$ and D. Belitz$^{2}$}
\affiliation{$^{1}$Institute for Physical Science and Technology and Department
                   of Physics, University of Maryland, College Park, MD 20742\\
         $^{2}$Department of Physics and Theoretical Science Institute, University
                of Oregon, Eugene, OR 97403}
\date{\today}
\begin{abstract}
It is shown that density-of-states fluctuations, which can be interpreted as the order-parameter
susceptibility $\chi_{\text{OP}}$ in a Fermi liquid, are anomalously strong as a result of the 
existence of Goldstone modes and associated strong fluctuations. In a 2-$d$ system with a 
long-range Coulomb interaction, a suitably defined $\chi_{\text{OP}}$ diverges as $1/T^2$ as a 
function of temperature in the limit of small wavenumber and frequency. In contrast, standard 
statistics suggest $\chi_{\text{OP}} = O(T)$, a discrepancy of three powers of $T$. The reasons 
behind this surprising prediction, as well as ways to observe it, are discussed.
\end{abstract}
\maketitle
The ordered phase in a classical Heisenberg ferromagnet can be understood as a stable fixed point (FP)
in a renormalization-group (RG) framework \cite{Wilson_Kogut_1974, Ma_1976}. Analogously,
the Fermi-liquid (FL) state in a many-electron system can be understood as an ordered 
state. This notion was pioneered by Wegner in the context of disordered systems \cite{Wegner_1979}.
More recently it was realized that the principle behind it is much more general and powerful, 
and can be applied to  clean FLs as well \cite{Belitz_Kirkpatrick_1997, Belitz_Kirkpatrick_2012}. 
An important feature of the resulting effective field theory \cite{Belitz_Kirkpatrick_2012} is that 
it integrates out all massive degrees of freedom to arrive at an effective soft-mode theory that
allows for a RG analysis \cite{other_theories_footnote}. As we will show in this Letter, it allows 
in particular for an analysis of the FL FP, associated corrections to scaling, 
and the scaling behavior of various observables. Within this
framework, the order parameter (OP) is the density of states (DOS) \cite{OP_footnote}, the field 
conjugate to the OP is a frequency-dependent chemical potential, and the spontaneously broken 
continuous symmetry is a rotational symmetry in frequency space that can be understood as the 
symmetry between retarded and advanced degrees of freedom. These quantities 
are analogous to the magnetization, the magnetic field, and the spin-rotational
symmetry, respectively, in a ferromagnet. The resulting Goldstone modes 
are the soft particle-hole excitations, i.e., four-fermion correlation functions that mix
advanced and retarded degrees of freedom. Their frequency $\omega$ scales linearly with the wave
number $k$, and they are responsible, {\it inter alia}, for the characteristic $\vert\omega\vert/k$
dependence of the Lindhard function \cite{Pines_Nozieres_1989}. They correspond to the
magnons in the ferromagnetic analogy; however, in contrast to magnons they are soft only at 
zero temperature ($T=0$).

Let us recall the behavior of the DOS in a FL as a function of the frequency $\omega$ (or the energy 
distance from the Fermi surface). We will focus on two-dimensional ($2$-$d$) systems, and will 
consider both a long-ranged Coulomb interaction and a short-ranged interaction. Results for general
dimensions $d>1$ will be reported elsewhere \cite{us_tbp}. For a Coulomb interaction, it is known 
that the DOS in $d=2$ is a nonanalytic function of $\omega$, namely,
\be
N(\omega) = 2\NF\left(1 + a\,\vert\omega\vert/\epsilonF\right) + o(\omega)
\label{eq:1}
\ee
Here $\NF$ is the DOS per spin at the Fermi level $\epsilonF$, and $a$ is a 
coefficient of $O(1)$. $o(\omega)$ denotes terms that vanish faster than linearly for 
$\omega\to 0$. This result has been derived within many-body perturbation theory, 
which yields a weak-coupling value for $a$, viz., $a = 1/4$ 
\cite{Khveshchenko_Reizer_1998, Mishchenko_Andreev_2002, LR_SR_footnote}. 
As we will show below, the soft-mode theory developed in 
Ref.\ \onlinecite{Belitz_Kirkpatrick_2012} establishes the 
$\vert\omega\vert$ as the {\em exact} leading frequency dependence, which is something 
that perturbation theory cannot achieve.  Technically, we will show that the frequency correction
in Eq. (\ref{eq:1}) is the leading correction to scaling at the FL FP. For the prefactor, to leading 
order in a loop expansion we recover the result from perturbation theory, $a=1/4$; this will 
acquire corrections if one goes to higher-loop order.
The linear frequency dependence is also consistent with simple scaling considerations,
which predict $N(\omega) - \NF \propto \vert\omega\vert^{d-1}$ \cite{Belitz_Kirkpatrick_2012},
and it constitutes a pseudogap as originally defined by Mott \cite{Mott_1968}. It is worth noting 
that this pseudogap is an intrinsic feature of {\em any} $2$-$d$ Fermi liquid. 
It has been observed in high-mobility tunnel junctions \cite{tunnel_junction_footnote}.

Given these results, the behavior of the OP susceptibility is of great interest. 
This quantity is an important measure of the strength of fluctuations
in the ordered phase, which provides a measure of the likelihood that the
ordered phase will become unstable as a function of a suitable parameter. In the case
of the classical Heisenberg ferromagnet, the OP susceptibility is
the magnetic susceptibility $\chi_{\text{m}} = \partial m/\partial h$. As a function
of the wave number $k$, the longitudinal component of $\chi_{\text{m}}$ diverges for 
$k\to 0$ for all dimensions $d<4$: $\chi_{\text{m,L}}(k) \propto 1/k^{4-d}$ \cite{Brezin_Wallace_1973}.
This reflects the strong fluctuations in the ordered phase due to the existence of magnons 
that couple to the OP. It is then natural to ask whether an analogous susceptibility
exists in the FL, and what its behavior at small wave numbers and frequencies is.

To answer this question, we represent the DOS as
\be
N(\omega) = \Re\,G({\bm x}=0,i\omega_n \to \omega + i0)/\pi
\label{eq:2}
\ee
in terms of the exact Green's function
\be
G({\bm x}=0,i\omega_n) = \langle {\rho}({\bm x},i\omega_n) \rangle\ ,
\label{eq:3}
\ee
with ${\rho}({\bm x},i\omega_n) = {\bar\psi}_n({\bm x})\,\psi_n({\bm x})$
in terms of fermion fields $\bar\psi_{n} ({\bm x})$ and $\psi_{n}({\bm x})$ that depend
on a fermionic Matsubara frequency $\omega_n = 2\pi T(n+1/2)$, 
($n = 0,\pm 1, \ldots$) \cite{density_footnote}.
We suppress the spin degree of freedom, which is not important for our purposes. 
The temporal Fourier transform is defined as
\be
\psi_n({\bm x}) = T^{1/2} \int_0^{1/T} d\tau\ e^{i\omega_n\tau}\,\psi({\bm x},\tau)\ ,
\label{eq:4}
\ee
with $\tau$ the imaginary-time variable. 
Now we consider the spatial Fourier
transform ${\rho}({\bm k},i\omega_n) = \int d{\bm x}\ e^{-i{\bm k}\cdot{\bm x}}\,{\rho}({\bm x},i\omega_n)$, and
the fluctuation $\delta{\rho} = {\rho} - \langle{\rho}\rangle$,
and define a correlation function
\be
\chi_{\text{OP}}({\bm k},i\omega_{n}) = \left\langle\delta{\rho}({\bm k},i\omega_n)\,\delta{\rho}(-{\bm k},i\omega_n)
   \right\rangle\ .
\label{eq:5}
\ee
$\chi_{\text{OP}}$ is the OP
susceptibility of interest. It describes both the response of the OP (i.e., the DOS) to its 
conjugate field, and its spontaneous fluctuations. The former is related to the latter by the
fluctuation-dissipation theorem. 

We first use simple statistical arguments to determine the behavior of $\chi_{\text{OP}}$ one
would expect in the absence of anomalous fluctuations. This will also make a general
point that will be important later. Consider
$\varphi_n({\bm x}) = \psi_n({\bm x})/\sqrt{T}$, and define the ``volume'' $V_T \equiv 1/T$ in the imaginary-time direction of
the space-time of quantum statistical mechanics. Then $\langle p_n({\bm x})\rangle \equiv
\langle{\bar\varphi}_n({\bm x})\,\varphi_n({\bm x})\rangle \propto V_T$ is a ``time-extensive''
quantity that is proportional to $V_T \equiv 1/T$. Now consider the relative fluctuation
$\langle(\delta p_n({\bm x}))^2\rangle/\langle p_n({\bm x})\rangle^2 \propto V_T/V_T^2 = 1/V_T = T$.
This just says that the relative fluctuation is proportional to $1/V_T$, as one would
expect from ordinary statistics. This yields an estimate for the fluctuations
of ${\rho}$:
\be
\langle(\delta{\rho}_n({\bm x},i\omega_n))^2\rangle \propto \frac{\langle(\delta p_n)^2\rangle}{V_T^2}
\propto \frac{\langle(\delta p_n)^2\rangle/V_T^2}{(\langle p_n\rangle/V_T)^2}
\propto 1/V_T = T\ .
\label{eq:6}
\ee
These arguments assume that there are no strong fluctuations in the
system that invalidate the simple statistics. Independent of this assumption, however, they
show that $\langle(\delta{\rho}_n({\bm x},i\omega_n))^2\rangle$, and hence $\chi_{\text{OP}}$, must
carry a factor of $T = 1/V_T$ that is the inverse linear extension of the imaginary-time axis.
We thus have
\be
\chi_{\text{OP}}({\bm k},i\omega_n;T) = T\,\theta({\bm k},i\omega_n;T)\ .
\label{eq:7}
\ee
In the absence of anomalous fluctuations, $\theta$ will scale as the zeroth power of
the wave number, the frequency, or the temperature; i.e., $\theta \sim 1$.

The last conclusion is not correct in an actual FL. To see this, we first give a
phenomenological scaling argument; later we show how to derive scaling from 
the effective field theory of Ref.\ \onlinecite{Belitz_Kirkpatrick_2012}. We first
consider a short-ranged interaction, in which case there is only one dynamical 
scaling exponent $z=1$. This reflects the fact that in the Goldstone modes of
the FL, the frequency scales as the wave number, $\omega \sim k$. The
DOS correction $\delta N = N - 2\NF$, which is the scaling part of the OP density, is
dimensionally an inverse volume times an inverse energy. In $d=2$, and with $z=1$,
it thus is expected to scale as an inverse length. The scaling assumption therefore is
that $\delta N$ obeys the homogeneity law
\be
\delta N(T) = b^{-1}\,\delta N(Tb)\ .
\label{eq:9}
\ee
This immediately yields $\delta N(T)  \propto T$, in agreement with Eq.\ (\ref{eq:1}). 
To obtain the behavior of the OP susceptibility, we remember that the field $h$
conjugate to the OP is a generalized chemical potential, which has a scale dimension
$[h] = z$. Adding $h$ as an argument of $\delta N$ in Eq.\ (\ref{eq:9}), differentiating
with respect to $h$, and adding the wave number argument yields for
$\chi_{\text{OP}} = \partial\,\delta N/\partial h$
\be
\chi_{\text{OP}}(T,k) = \chi_{\text{OP}}(Tb,kb) = f_{\chi}(T/k)\ ,
\label{eq:10}
\ee
with $f_{\chi}$ a scaling function.
If we combine this with the conclusion from Eq.\ (\ref{eq:7}) that $\chi_{\text{OP}}$
must be proportional to $T$ for fundamental quantum statistical reasons, we find
\be
\chi_{\text{OP}}(T,k) \propto \begin{cases} T/k & \text{for $T\to 0$ at fixed $k$} \\
                                                                    T/T = O(1) & \text{for $k\to 0$ at fixed $T$\ .}
                                              \end{cases} 
\label{eq:11}
\ee
This result is not consistent with the naive statistical arguments given above: If the DOS were 
normally distributed, we would have $\chi_{\text{OP}} \propto 1/V_T$ and $\theta \sim 1$.
Instead, we see that the quantity $\theta$ in Eq.\ (\ref{eq:7}) scales as $\theta \sim 1/k \sim 1/T$,
and $\chi_{\text{OP}} \sim 1$. This divergence of the relative DOS fluctuations reflects
the strong fluctuations in the system that are a consequence of the existence of the Goldstone
modes. As we will see below, a long-ranged Coulomb interaction further amplifies these effects.

Before we turn to the case of a Coulomb interaction, let us show how
these results can be derived without invoking a scaling assumption, by performing a
RG analysis of the effective field theory of Ref. \ \onlinecite{Belitz_Kirkpatrick_2012}.
We note that even though the FL FP is not a critical FP, it nevertheless displays scale
invariance due to the existence of Goldstone modes. Therefore, very useful results for 
the entire FL phase can be obtained from very simple RG arguments.

The theory is formulated in term of a soft matrix field $q_{nm}({\bm k})$ 
\cite{qslash_footnote, density_formulation_footnote}, which encodes the soft components of 
bilinear fermion fields $\bar\psi_n \psi_m$, viz., those products with $nm<0$. Their softness is 
guaranteed by a Ward identity. The effective action ${\cal A}$ takes the form
of an expansion in powers of $q$, see Eqs.\ (4.51) in Ref.\ \onlinecite{Belitz_Kirkpatrick_2012}.
In a symbolic notation that shows only quantities that carry a scale dimension, viz., the fields
$q_{nm}({\bm k}) \equiv q$, and factors of volume $V$, wave number $k$, and frequency $\omega$
(which we do not need to distinguish from factors of temperature for our purposes), it takes the form
\bea
{\cal A} &=& \frac{1}{V}\sum_{k,\omega} [k + \omega + \gamma\omega]\, q^2 
                  + \frac{c_3}{V^2}\sum_{\{k,\omega\}} [\omega + O(\omega^3)]\, q^3
\nonumber\\ 
&& \hskip -20pt+ \frac{c_4}{V^3}\sum_{\{k,\omega\}} [k + \omega + \omega^2/k + O(\omega^3)]\, q^4
     + O(q^5).
\label{eq:12}
\eea
Here the sums are over the appropriate sets of wave vectors and frequencies, and the powers of $k$ and
$\omega$ in each term follow from the behavior of the convolutions of Green's functions that make up the
vertices of the theory in the limit of long wavelengths and small frequencies, 
see Ref.\ \onlinecite{Belitz_Kirkpatrick_2012}. As mentioned above, $\omega$ can stand for either
frequency or temperature, and $\gamma$ represents the interaction amplitude. $c_3$ and $c_4$ are 
schematic coupling constants; $c_3 \propto \gamma$.
We now determine the FP action that describes the FL. We use Ma's method of choosing scale
dimensions for all relevant quantities and then showing self-consistently that these choices
lead to a stable FP \cite{Ma_1976}. We assign a scale dimension $[k]=1$ to wave numbers, and $[\omega]=1$
to frequencies (i.e., we choose a dynamical exponent $z=1$). The latter choice reflects the 
linear dispersion relation of the soft modes, see
the first term in Eq.\ (\ref{eq:12}), which in a FL we do not expect to be changed by 
renormalization. We further do not expect the power of wave number (or frequency) in the
Gaussian vertex to be renormalized, and therefore assign a scale dimension $[q({\bm k})] = -(d+1)/2$
and $[q({\bm x})] = -(d-1)/2$ to the field as a function of ${\bm k}$ and ${\bm x}$,
respectively (i.e., we choose the exponent $\eta$ to be zero.) With these choices,
the $q^2$ term in Eq.\ (\ref{eq:12}) is dimensionless; in particular, $[\gamma]=0$. For the cubic 
term we have $[c_3] = -(d-1)/2$, for the quartic one, $[c_4] = -(d-1)$, etc. Each additional
power of $q$ reduces the scale dimension of the
corresponding coupling constant by $-(d-1)/2$. The FP action is thus given by the Gaussian
term alone, and all terms of higher order in $q$ are irrelevant with respect to the FL FP in
all dimensions $d>1$. 
It follows by standard arguments \cite{Wilson_Kogut_1974} that this remains true order by
order in a loop expansion. All coefficients will in general acquire finite renormalizations, but
the structure of the theory will not change. An important ingredient in this chain of arguments
is the Ward identity proven in Ref.\ \onlinecite{Belitz_Kirkpatrick_2012}, which identifies $q$
as a soft mode. This assures that the vertices in Eq.\ (\ref{eq:12}) will remain soft under
renormalization.

We now use the above conclusions to determine the observables we are interested in. Let us
first consider the DOS, Eqs.\ (\ref{eq:2}, \ref{eq:3}). It is given as an expectation value of
${\bar\psi}_n \psi_n$, which is a massive mode. However, it couples to the soft modes
and hence can be expressed as a series of $q$-correlation functions \cite{NLsM_footnote}. 
Schematically, 
\be
N \sim 1 + \langle q^2\rangle + \langle q^4\rangle + \ldots
\label{eq:12'}
\ee
The RG arguments given above guarantee that the leading contribution to the DOS
correction is given by the term quadratic in $q$. For the scale dimension of the leading scaling part of $\delta N$ this implies
$[\delta N] = 2[q({\bm x})] = d-1$. For $d=2$ this yields Eq.\ (\ref{eq:9}). By an analogous
argument we find $[\chi_{\text{OP}}] = d-1-z = d-2$, which for $d=2$ yields Eq.\ (\ref{eq:10}).
We thus have derived scaling from the field theory via a RG treatment.

In order to determine the correct scaling behavior in  the case of a Coulomb interaction, an 
explicit calculation is needed in addition to general arguments because of the presence of a 
dangerous irrelevant variable (DIV). This is analogous to the case of a classical ferromagnet
in $d>4$, where one needs to explicitly calculate the equation of state to understand why
hyperscaling breaks down \cite{Ma_1976}. To study this case, we replace the constant 
interaction amplitude $\gamma$ by the dynamically screened Coulomb potential.  In $d=2$, 
the latter has the schematic structure
\be
U(k,i\omega) = \frac{1}{k + \kappa - \frac{\kappa\omega/\vF k}{\sqrt{1+\omega^2/(\vF k)^2}}}
\approx \frac{1}{uk + \kappa(\vF k)^2/\omega^2}\ .
\label{eq:13}
\ee
Here $\kappa$ is the screening wave number, and $\vF$ is the Fermi velocity. The second expression 
is valid in the limit $\vF k \ll \omega$, and we have introduced a coupling constant $u$ whose bare
value is equal to 1. In the denominator we recognize the plasmon excitation, with its
characteristic $\omega \sim k^{1/2}$ scaling. We also see the well-known fact that screening 
breaks down at nonzero frequencies. Counting powers again, we see that $u$ is irrelevant
with scale dimension $[u]=-1$. However, it turns out that $u$ is a DIV in $d=2$ with respect to 
$\chi_{\text{OP}}$ (but not to the DOS). 

To demonstrate this we first calculate the $\langle q^2\rangle$
term in Eq.\ (\ref{eq:12'}), which is the leading contribution to $\delta N$. Using the 
formalism of Ref.\ \onlinecite{Belitz_Kirkpatrick_2012} we find that it takes the form of
a frequency-momentum integral over a vertex function $V(k,i\omega)$ times $U(k,i\omega)$.
The former has the structure $V(k,i\omega) = v(\omega/\vF k)/k^2$, with $v(x\to\infty) \propto 1/x^2$. 
The integral that represents the DOS correction then has the structure \cite{DOS_footnote}
\be
\delta N \sim \int_0^{\Lambda} dk\,k \int_T^{\vF\Lambda} d\omega\ \frac{1}{k^2}\,v(\omega/\vF k)\,U(k,i\omega)\ ,
\label{eq:14}
\ee  
where $\Lambda$ is an ultraviolet momentum cutoff. From Eq.\ (\ref{eq:14}) we see by power
counting that $N(T) \propto \text{const.} - T$, in agreement with Eqs.\ (\ref{eq:9}),
(\ref{eq:1}). This is true for both the long-range and the short-range cases (the latter is
recovered by replacing $U$ with a constant); the singular
nature of the screened Coulomb interaction at $\omega\neq 0$ does not suffice to change
the behavior in $d=2$ \cite{d_less_than_2_footnote}.       

Now consider the OP susceptibility for $d=2$. It follows from Eq.\ (\ref{eq:5}) that $\chi_{\text{OP}}$
is given by an integral analogous to the one in Eq.\ (\ref{eq:14}) with the integrand squared 
and an extra factor of $T$. An external wave number effectively serves as a lower cutoff
on the wave-number integral, and we have schematically
\be
\chi_{\text{OP}}(T,k) \sim T \int_k^{\Lambda} \hskip -5pt dp\,p \int_T^{\vF\Lambda}\hskip -12pt d\omega\ \left(
     \frac{1}{p^2}\,v(\omega/\vF p)\,U(p,i\omega)\right)^2\,.
\label{eq:15}
\ee   
The prefactor of $T$ is just the trivial ``volume'' factor $1/V_T$ discussed in the context
of Eqs.\ (\ref{eq:6}, \ref{eq:7}). In the short-range case, it is easy to see that Eq.\ (\ref{eq:15})
yields Eq.\ (\ref{eq:11}) in the respective limits. In the long-range case, using Eq.\ (\ref{eq:13}) 
in Eq.\ (\ref{eq:15}) it is easy to see that the integral diverges as $1/u^{3/2}$ for $u\to 0$.
Restoring the prefactors, we find 
\bse
\label{eqs:16}
\be
\chi_{\text{OP}}(k) = \frac{\kappa^2}{90\pi^2}\,\frac{T}{T^3}\,\ln(T^2/\vF^2\kappa k)
\label{eq:16a}
\ee
This result is valid for $T^2/\vF k \gg \vF\kappa$, and to logarithmic accuracy. In the 
opposite limit, we find \cite{prefactor_footnote}
\be
\chi_{\text{OP}}(k) \propto \frac{\kappa^{1/2}}{\vF^3}\,\frac{T}{k^{3/2}}\ ,
\label{eq:16b}
\ee
Finally, for $T\to 0$ and nonzero external frequency $\omega$ one finds, to logarithmic 
accuracy for $\omega^2\gg \vF^2\kappa k$,
\be
\chi_{\text{OP}}(k,i\omega) = \frac{8}{3\pi^2}\,\frac{\kappa^2 T}{\omega^3}\,
   \ln(\omega^2/\vF^2\kappa k)
\label{eq:16c}
\ee
\ese
We see that the OP susceptibility, normalized to account for a trivial factor of $T$,
Eq.\ (\ref{eq:7}), diverges in the limit of vanishing wave number $k$ as $1/k^{3/2}$,
or as $1/T^3$ or $1/\omega^3$ in the limits of vanishing
temperature or frequency, respectively. This is a very strong effect; uncorrelated
statistics would lead to a constant for the same quantity.
Comparing Eqs. (\ref{eq:16b}) and (\ref{eq:16a}, {\ref{eq:16c}) we see that $T$ or $\omega$
scale as $k^{1/2}$, which reflects the integral being dominated by the plasmon
time scale. Although the latter is subleading by power counting, it dominates the 
scaling due to the DIV $u$. 
Notice that $\chi_{\text{OP}}(k\to 0)$ diverges even at a
$T>0$ since the plasmon, as a density fluctuation governed by
a conservation law, remains soft even at $T>0$.

Also of interest is the homogeneous susceptibility in a finite system of
linear dimension $L$. In that case, $k$ in Eqs.\ (\ref{eqs:16}) gets replaced by 
$1/L$ \cite{us_tbp}.

We finally add some comments about the experimental relevance of the above results.
The spectrum of the local Green's function, Eq.\ (\ref{eq:3}), is what is often referred
to as the local density of states (LDOS) \cite{Chan_Heller_1997, LDOS_footnote}. The
LDOS gives the dominant contribution to the tunneling current in a scanning tunneling
experiment \cite{Tersoff_Hamann_1985}; its spatial average, which is the DOS, is
measured in a classic tunnel junction. Our OP susceptibility, Eq. (\ref{eq:5}), describes
the spatially averaged fluctuations of the LDOS. A two-tip tunneling experiment
\cite{Niu_Chang_Shih_1995, Chan_Heller_1997} should be able to give information
about this quantity.

We are grateful to S. Gregory and H. Manoharan for helpful discussions.
This work was supported by the National Science Foundation under Grant Nos.
DMR-09-29966, DMR-09-01907, and PHY-10-66293. Part of this work was
performed at the Aspen Center for Physics.


\begin{thebibliography}{29}
\bibitem[{OP_()}]{OP_footnote}
\bibinfo{note}{More precisely, the OP is the single-particle spectral density,
  whose zeroth wave-number moment is the DOS. For simplicity, we refer to the
  DOS as the OP.}
\bibitem[{den({\natexlab{a}})}]{density_footnote}
\bibinfo{note}{${\rho}({\bm x},i\omega_n)$ can be interpreted as the density of
  particles at point ${\bm x}$ and (complex) energy $i\omega_n$. Note, however,
  that $\rho$ is local in frequency space, since it is defined as a product of
  two fermion fields at the same frequency $\omega_n$, whereas the actual
  particle number density is local in time and involves a convolution in
  frequency space.}
\bibitem[{qsl()}]{qslash_footnote}
\bibinfo{note}{The theory also contains a second field $\qslash_{nm}({\bm k})$.
  However, this has the same scale dimension as $q$, and for our purposes we do
  not need to distinguish between $q$ and $\qslash$.}
\bibitem[{pre()}]{prefactor_footnote}
\bibinfo{note}{The prefactor in Eq.\ (\ref{eq:16b}) is difficult to determine
  since the wave number integral involves a convolution.}
\end{thebibliography}

\end{document}